\documentclass[prx,twocolumn,floatfix,superscriptaddress,longbibliography]{revtex4-2}
\usepackage{amssymb,amsmath,amstext}
\usepackage[algo2e,ruled,longend]{algorithm2e}
\usepackage{graphicx}
\usepackage{subfigure}
\usepackage{epstopdf}
\usepackage{color}
\usepackage{xcolor}
\usepackage{bm}
\usepackage{appendix}
\usepackage[T1]{fontenc}
\usepackage{bbold}
\usepackage{bbm}
\usepackage{physics}
\usepackage{latexsym}
\usepackage[english]{babel}
\usepackage[colorlinks=true,citecolor=blue,linkcolor=magenta]{hyperref}

\newcommand{\beq}{\begin{equation}}
\newcommand{\eeq}{\end{equation}}
\newcommand{\beqnn}{\begin{equation*}}
\newcommand{\eeqnn}{\end{equation*}}
\newcommand{\bea}{\begin{eqnarray}}
\newcommand{\eea}{\end{eqnarray}}
\newcommand{\beann}{\begin{eqnarray*}}
\newcommand{\eeann}{\end{eqnarray*}}
\newcommand{\bes} {\begin{subequations}}
\newcommand{\ees} {\end{subequations}}

\begin{document}

\title{Diabatic Quantum Annealing for the Frustrated Ring Model}

\author{Jeremy C\^ot\'e}
\affiliation{Theoretical Division, Los Alamos National Laboratory, Los Alamos, New Mexico 87545, USA}
\affiliation{Institut quantique \& D\'{e}partement de physique, Universit\'{e} de Sherbrooke, Sherbrooke, Qu\'{e}bec J1K 2R1, Canada}

\author{Fr\'ed\'eric Sauvage}
\affiliation{Theoretical Division, Los Alamos National Laboratory, Los Alamos, New Mexico 87545, USA}

\author{Mart\'in Larocca}
\affiliation{Theoretical Division, Los Alamos National Laboratory, Los Alamos, New Mexico 87545, USA}
\affiliation{Center for Nonlinear Studies, Los Alamos National Laboratory, Los Alamos, New Mexico 87545, USA}

\author{Mat\'ias Jonsson}
\affiliation{Theoretical Division, Los Alamos National Laboratory, Los Alamos, New Mexico 87545, USA}
\affiliation{Department of Physics, Carnegie Mellon University. Pittsburgh, Pennsylvania 15217, USA}

\author{Lukasz Cincio}
\affiliation{Theoretical Division, Los Alamos National Laboratory, Los Alamos, New Mexico 87545, USA}

\author{Tameem Albash}
\affiliation{Department of Electrical and Computer Engineering, University of New Mexico, Albuquerque, New Mexico
87131, USA}
\affiliation{Department of Physics and Astronomy and Center for Quantum Information and Control, University of New
Mexico, Albuquerque, New Mexico 87131, USA}

\begin{abstract}
Quantum annealing is a continuous-time heuristic quantum algorithm for solving or approximately solving classical optimization problems. The algorithm uses a schedule to interpolate between a driver Hamiltonian with an easy-to-prepare ground state and a problem Hamiltonian whose ground state encodes solutions to an optimization problem. The standard implementation relies on the evolution being adiabatic: keeping the system in the instantaneous ground state with high probability and requiring a time scale inversely related to the minimum energy gap between the instantaneous ground and excited states. However, adiabatic evolution can lead to evolution times that scale exponentially with the system size, even for computationally simple problems. Here, we study whether non-adiabatic evolutions with optimized annealing schedules can bypass this exponential slowdown for one such class of problems called the frustrated ring model. For sufficiently optimized annealing schedules and system sizes of up to 39 qubits, we provide numerical evidence that we can avoid the exponential slowdown. Our work highlights the potential of highly-controllable quantum annealing to circumvent bottlenecks associated with the standard implementation of quantum annealing.
\end{abstract}
\maketitle

\section{Introduction} 
\label{sec:Introduction}

The ubiquity of optimization problems across disciplines and the incredible computational resources they consume continues to motivate researchers to explore new and more efficient approaches to tackle them. Though quantum computing has so far offered limited provable computational advantages for combinatorial optimization problems \cite{Gro1997,Bra2000,Cer2000,Rol2002,Chi2003,Amb2004,Wie2012,Som2012,Mon2015,Bra2016,Man2016,Amb2017,Cam2019,Bra2022}, we expect that the library of quantum approaches for tackling different classes of optimization problems will expand as quantum information processors mature and become more readily available.

Quantum annealing (QA) is a generic quantum approach to tackling combinatorial optimization problems. Starting from an efficiently prepared ground state of an initial Hamiltonian $\hat{H}_d$, the system is evolved according to an interpolating Hamiltonian between $\hat{H}_d$ and a problem Hamiltonian $\hat{H}_p$, which ground state encodes the solution to the optimization problem. At the end of the evolution, the algorithm is deemed successful if the quantum state has high overlap with the ground state of $\hat{H}_p$.

A sufficient but not necessary condition for QA to succeed is for the evolution to satisfy the adiabatic condition, which requires the total evolution time $T$, also called the annealing time, to scale as an inverse power of the minimum gap encountered along the interpolations~\cite{Bor1928,Kat1950,Jan2007}. When it does, QA is sometimes referred to as quantum adiabatic optimization (QAO)~\cite{Kad1998,Far2001}, and it belongs to the adiabatic paradigm of quantum computing~\cite{Far2000,Aha2007}. The adiabatic theorem of quantum mechanics~\cite{Bor1928,Kat1950,Jan2007} provides a guarantee that if the evolution is sufficiently slow, then the state at the end of the evolution will have high overlap with the ground state of $\hat{H}_p$. The efficiency of the algorithm is thus given by the scaling of the minimum gap encountered along the interpolating schedule with the system size.

However, enforcing adiabaticity for all realizations of QA can result in exponentially slow evolutions even for simple problems~\cite{van2001,Rei2004}. Furthermore, the only provable exponential speedups currently known for QA involves an evolution that is \emph{not} adiabatic~\cite{Som2012}. This suggests that it might be meaningful to consider heuristic continuous-time quantum optimization algorithms that are not necessarily adiabatic, and for which we can design the interpolation to take advantage of non-adiabatic transitions. This approach is sometimes called `diabatic quantum annealing' (DQA)~\cite{Cro2021}. The added flexiblity of DQA comes at the cost of identifying a suitable interpolation schedule, which is a non-trivial optimization problem in itself. If our goal is to search for interpolating schedules that improve on the evolution times of the adiabatic approach, we must ensure that the computational resources to identify these schedules also scale favorably with the system size.

In this work, we consider the performance of DQA on a simple class of optimization problems, known to be exponentially hard for QAO~\cite{Rob2020}. The problem is a one-dimensional frustrated ring model whose ground state can be found analytically, but it poses a challenge for standard implementations of QAO because it exhibits a perturbative crossing~\cite{Ami2009}, leading to exponentially slow evolution times. We will show that we can identify interpolating schedules that solve the optimization problem more efficiently than QAO.

The paper is organized as follows. In Sec.~\ref{sec:Model}, we review the frustrated ring model and why it exhibits an exponentially closing gap with the system size. In Sec.~\ref{sec:methods}, we present our algorithm for optimizing the annealing schedules and annealing times. In Sec.~\ref{sec:Results}, we present our numerical results for our optimized schedules and the performance of the DQA algorithm. We conclude and highlight some future directions in Sec.~\ref{sec:Conclusion}.

\section{Model}
\label{sec:Model}

\begin{figure}[ht]
    \centering
    \includegraphics[width=0.45\textwidth]{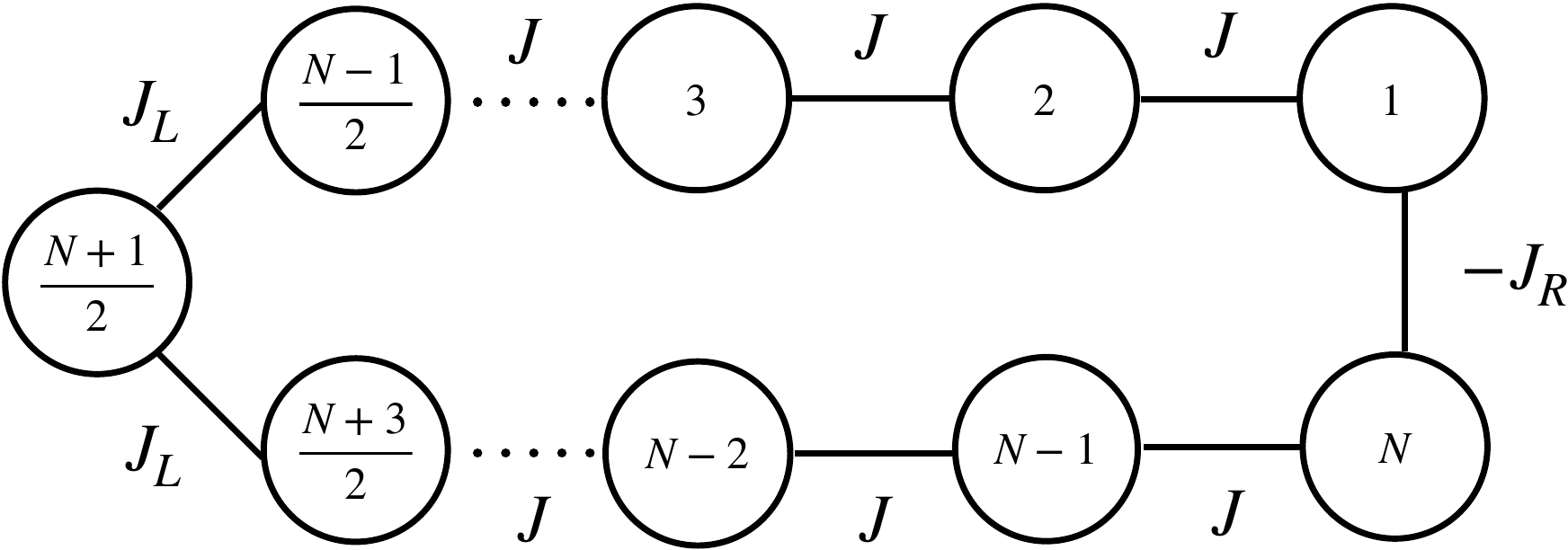}
    \caption{A visualization of the frustrated ring model with a total of $N$ (odd) spins with our indexing convention for the sites.}
    \label{fig:model}
\end{figure}

For our problem Hamiltonian, we focus on the frustrated ring model of Ref.~\cite{Rob2020} defined on $N$ Ising spins arranged in a ring. The problem Hamiltonian is given by:
\begin{equation}
    \label{eq:Hamiltonian}
    \hat{H}_p = - \sum_{j = 1}^N J_{j} \hat{Z}_j \hat{Z}_{j+1},
\end{equation}
where $J_{j}$ denotes the coupling between spins $j$ and $j+1$, $\hat{Z}_j$ is the Pauli operator $\hat{Z}$ acting on spin $j$, and $\hat{Z}_{N+1} \equiv \hat{Z}_1$. We denote the $+1$ eigenstate of $\hat{Z}$ by $\ket{0}$, and the $-1$ eigenstate by $\ket{1}$. For simplicity, we restrict ourselves to the case where $N$ is odd. There are three types of couplings, and using the indexing convention in Fig.~\ref{fig:model}, they are:
\begin{equation}
    \label{eq:couplings}
    J_j =
        \left\{
		    \begin{array}{lll}
			    -J_R & \mbox{if } j = N, \\
			    J_L & \mbox{if } j =  \frac{1}{2} \left( N \pm 1 \right), \\
			    J & \mbox{otherwise.}
		    \end{array}
	    \right.
\end{equation}
For the rest of the paper, we take the couplings to satisfy $0< J_R < J_L < J = 1$. With these choices, the couplings associated with $J_L$ and $J$ are ferromagnetic, and the couplings associated with $J_R$ are antiferromagnetic. 

The doubly degenerate ground states of this Hamiltonian are given by the states $\ket{0}^{\otimes N}$ and $\ket{1}^{\otimes N}$, which satisfy all the ferromagnetic couplings but violate the one antiferromagnetic coupling of the Hamiltonian. The ground state energy is thus given by:
\begin{equation}
    \label{eq:ExactGSenergy}
    E_{0} = - \sum_{j = 1}^N J_j = - (N-3)J + J_R - 2J_L .
\end{equation}
On the other hand, the first excited states satisfy the antiferromagnetic coupling but violate one of the two $J_L$ ferromagnetic couplings, giving rise to a four-fold degenerate first excited eigenspace with energy:

\begin{equation}
    \label{eq:ExactExcitedEnergy}
    E_{1} = -(N-3)J - J_R =  E_{0} + 2\left(J_L - J_R \right) .
\end{equation}
Thus, the difference $J_L - J_R$ determines the ground state energy gap of the problem Hamiltonian.

We now consider a continuous-time algorithm with a time-dependent Hamiltonian $\hat{H}(t)$ that interpolates between a uniform transverse field Hamiltonian $\hat{H}_d = - \sum_{j=1}^N \hat{X}_j$, where $\hat{X}_j$ is the Pauli operator $\hat{X}$ acting on site $j$, and the problem Hamiltonian. That is,
\begin{equation}
    \label{eq:HpHd}
    \hat{H}(t) = \left(1-A(t)\right) \hat{H}_d + A(t) \hat{H}_p = \hat{H}_d + A(t) \left( \hat{H}_p - \hat{H}_d \right) \ ,
\end{equation}
where $A(t)$ is the annealing schedule of our algorithm. For the remainder of the paper, we  require $A(0) =0$ and $A(T) =1$, where $T$ is the total annealing time for the protocol. 

An important feature of this combination of problem Hamiltonian and mixer Hamiltonian is that it gives rise to a `perturbative crossing' \cite{Ami2009} in the energy spectrum. To see this, we first express the ground states and first excited states as symmetric and anti-symmetric combinations (for simplicity we write this for the case of $N=7$):
\bes \label{eqt:symmetricCombo}
\begin{align}
    \ket{\Phi_0^{(\pm)}} & = \frac{1}{\sqrt{2}} \left( \ket{0000000} \pm \ket{1111111} \right) \ , \\
    \ket{\Phi_1^{(\pm)}} & = \frac{1}{\sqrt{2}} \left( \ket{0001111} \pm \ket{1110000} \right) \ ,\\
     \ket{\Phi_{1'}^{(\pm)}} & = \frac{1}{\sqrt{2}} \left( \ket{0000111} \pm \ket{1111000} \right) \ . 
\end{align}
\ees
Because the Hamiltonian $\hat{H}(t)$ commutes with the operator $\hat{X}^{\otimes N}$, the energy eigenstates can be labeled by the two eigenvalues $\pm 1$ of this operator. We only need to concern ourselves with the $+1$ eigenvalue states (the reason for this will become apparent when we discuss the initial state of our continuous-time algorithm), which in turn means we only need to consider the symmetric combinations of states in Eq.~\eqref{eqt:symmetricCombo}.
The two first excited states $\Phi_1^{(+)}$ and $\Phi_{1'}^{(+)}$ are coupled by the transverse field, so the degeneracy of the first excited state of the problem Hamiltonian is broken when we turn on the parameter $1-A$. At first order in perturbation theory, the unique instantaneous first excited state has energy 
\beq
    E_1(A)  = E_1 A - (1-A) \ .
\eeq
The ground state energy is corrected to $ E_0(A)  = (1-A)E_0$ at the same order.  As $A$ decreases from 1, first order perturbation theory predicts an energy level crossing at:

\beq \label{eqt:crossing}
A_\ast = \frac{1}{1 + E_1 - E_0} \ ,
\eeq
which will be corrected to an exponentially small (in $N$) avoided level crossing in the full quantum theory~\cite{Rob2020}, with Eq.~\eqref{eqt:crossing} being approximately the position of the minimum gap. For example, with $J_R = 0.45, J_L = 0.5$ we find $A_\ast \approx 0.9091$.
\section{Optimizing the Annealing Schedules}\label{sec:methods}

Starting from the state $\ket{\psi(0)} = \ket{+}^{\otimes N}$, which is the ground state of $\hat{H}(0)$, our objective is to prepare a state $\ket{\psi(T)} =\mathrm{Texp}\left( \int_0^T \hat{H}(t) dt \right) \ket{\psi(0)}$, with $\mathrm{Texp}$ being the time ordered exponential, that has a large overlap with the (relevant) ground state of $\hat{H}_p$. We provide details of how we simulate the dynamics in Appendix~\ref{sec:Simulation}. In order to avoid unfairly steering the optimization of the annealing schedule towards the known solution, we choose the cost function to minimize to be the energy $E(T) = \bra{\psi(T)} \hat{H_p} \ket{\psi(T)}$. We note that this choice is general and can be applied to any problem Hamiltonian. In order to determine the success of our optimization, we choose a threshold value $\Delta_E$ and consider any state $\ket{\psi(T)}$ with $E(T) \leq E_0 + \Delta_E$ to be a success. This choice does use our knowledge of the true ground state, and we use it to provide no ambiguity in the scaling performance of our optimization method.  In more realistic situations, where the ground state energy of the problem Hamiltonian is not known, one can use more sophisticated stopping criteria such as the optimal stopping discussed in Ref.~\cite{Vin2016}, where a further cost associated with every call to the algorithm is introduced. 

Our goal is to identify the lowest time $T$ and corresponding schedule which, given a target energy threshold value, ensures the success of the optimization. Other objectives besides the lowest time could be used, but we do not pursue them in this work. For example, if our objective is to minimize the total time to find a state with or below a target energy $E_0 + \Delta_E$, a more suitable metric might be the time-to-solution~\cite{Ron2014}, which balances spending more time per independent run of the algorithm for a higher success probability against performing multiple fast runs of the algorithm to boost a lower success probability.

As a baseline, we consider the linear annealing schedule $A(t) = t/T$, with $T$ large enough to satisfy the adiabatic condition. As we discussed in Sec.~\ref{sec:Model}, Hamiltonian $\hat{H}(t)$ exhibits along its interpolation a minimum energy gap between its first excited state and ground state that closes exponentially with the system size. This implies that an adiabatic protocol would require an evolution time that scales exponentially with the system size, such that any adiabatic protocol is exponentially slow in finding the ground state with high probability of such a computationally simple one-dimensional Ising problem.

\begin{figure}[ht]
    \centering
    \includegraphics[width=0.48\textwidth]{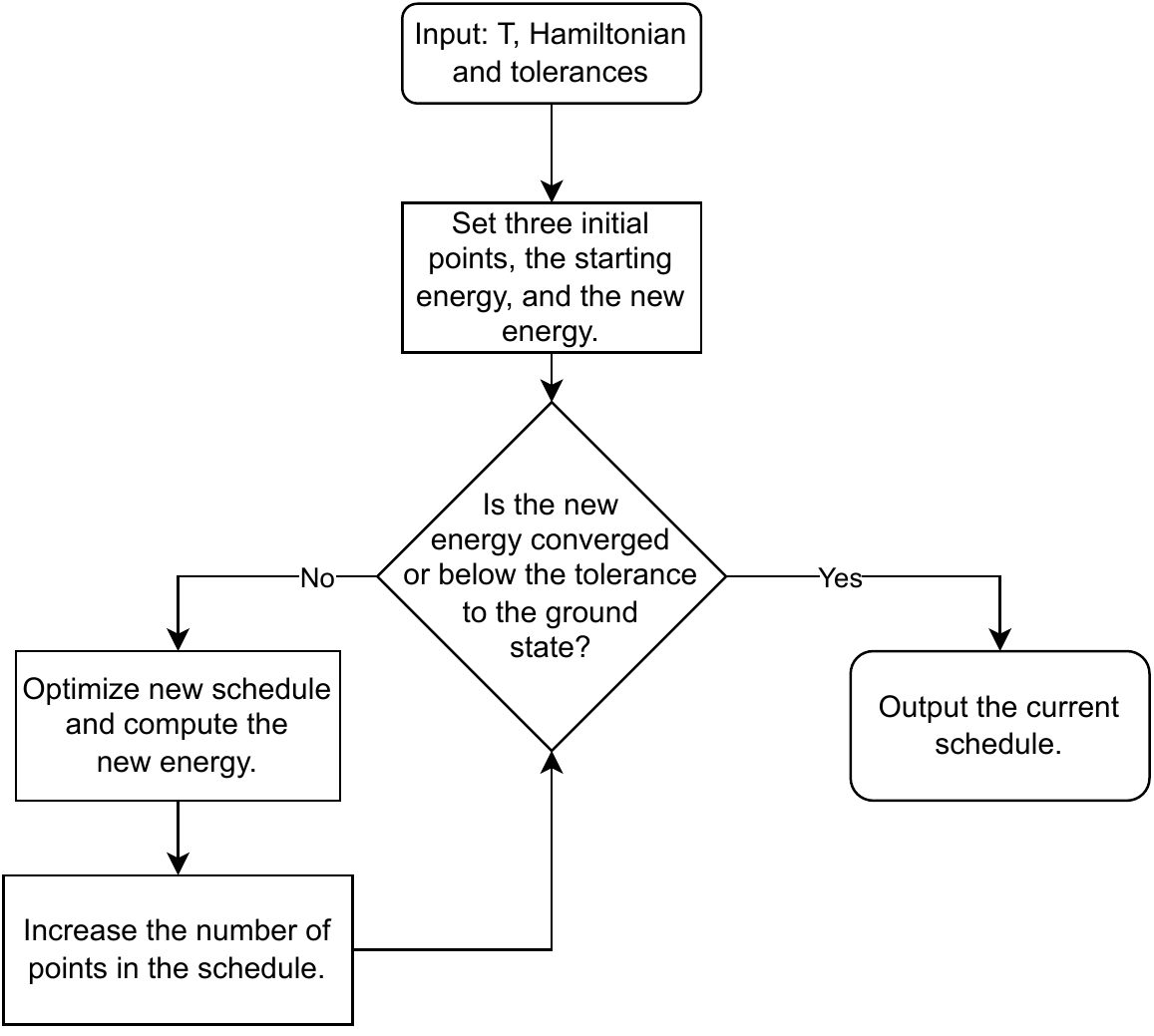}
    \caption{The inner routine of our optimization algorithm for optimizing the annealing schedule $A(t)$. It provides our reasoning behind Algorithm~\ref{alg:OptimizeA}.}
    \label{fig:optimizeSchedule}
\end{figure}

Our objective is to optimize the annealing time and schedule of DQA to avoid this exponential slowdown. To do so, we divide our optimization procedure into an inner and outer routine. Within the inner routine, which we visualize in Fig.~\ref{fig:optimizeSchedule}, the annealing time $T$ is fixed and we optimize the schedule $A(t)$ to minimize the energy of the prepared state. We discuss the details of this in Sec.~\ref{sec:OptimizingSchedule}.

In addition, the outer routine (Fig.~\ref{fig:optimizeT}) aims at minimizing the annealing time $T$ required to achieve a target threshold energy. We do this by iteratively decreasing (or increasing) $T$ depending on the success (or failure) of the inner routine. The algorithm repeats these steps until we identify a sufficiently narrow range of the values of $T$ for which the algorithm has found a successful schedule (Sec.~\ref{sec:OptimizingAnnealingTime}). We now describe the inner and outer routines in further detail.

\SetKwComment{Comment}{/* }{ */}

\begin{algorithm2e}
\caption{FindSchedule} \label{alg:OptimizeA}
\KwIn{$J_R$, $J_L$, $J$, $E_{0}$, $N$, $T$, $\delta t$, $\delta E, c$}
\KwOut{$A(t)$}
 $k \gets 3$\;
 $E_{\mathrm{previous}} \gets 0$\;
 $E(T) \gets \infty$\;
 \While{$E(T) - E_{0} \geq 2c(J_L - J_R)$ and $|E(T) - E_{\mathrm{previous}} | \geq \delta E$}{
    \If{$k > 3$}{
        $E_{\mathrm{previous}} \gets E(T)$\;
    }
    $p \gets \text{InitialPoints}(k)$  \tcp*[h]{Get $\left\{ a_j \right\}$}\;
    $A(t, p) \gets \mathrm{Optimize}(p)$\;
    $E(T) \gets \mathrm{ComputeEnergy}(A(t,p))$\;
    $k \gets 2k + 1$\;
}
$A(t) \gets A(t,p)$\;
\Return $A(t)$\;
\end{algorithm2e}

\subsection{Optimizing the schedule}
\label{sec:OptimizingSchedule}
The main routine of our algorithm involves optimizing the annealing schedule $A(t)$ such that we minimize $E(T) - E_{0}$ (Algorithm~\ref{alg:OptimizeA}), for a fixed time $T$. In particular, we want to ensure our output state $\ket{\psi(T)}$ has enough overlap with the true ground state $\ket{\Phi_0^+}$ so that we could in principle implement this schedule on a quantum processor and have a finite probability of measuring the ground states.

We parameterize the schedule $A(t)$ in terms of a finite number of continuous parameters corresponding to weights over a predefined set of functions. For instance, these functions could be a basis of trigonometric functions~\cite{Zho2020}, a basis of polynomials~\cite{Qui2019}, ``bang-bang'' or on-off pulses (standard in the quantum approximate optimization algorithm~\cite{Yan2017,Bra2021}), or others~\cite{Che2020}. In our case, we opt for a linear piecewise decomposition~\cite{Mat2021}, which we iteratively refine and now describe.

To parameterize $A(t)$, we begin by splitting our annealing interval $\left[0, T \right]$ into four equal intervals separated at $t_1 = T/4$, $t_2 = T/2$, and $t_3 = 3T/4$. We choose three random values $a_1$, $a_2$, and $a_3$ in the interval $\left[ 0, 1  \right]$, and then set $A(t_j) = a_j$. This defines the initial annealing schedule $A(t)$, with three linear functions connecting the three points (as well as the start and end points, $A(0) = 0$ and $A(T) = 1$).

\begin{figure}[ht]
    \centering
    \includegraphics[width=0.48\textwidth]{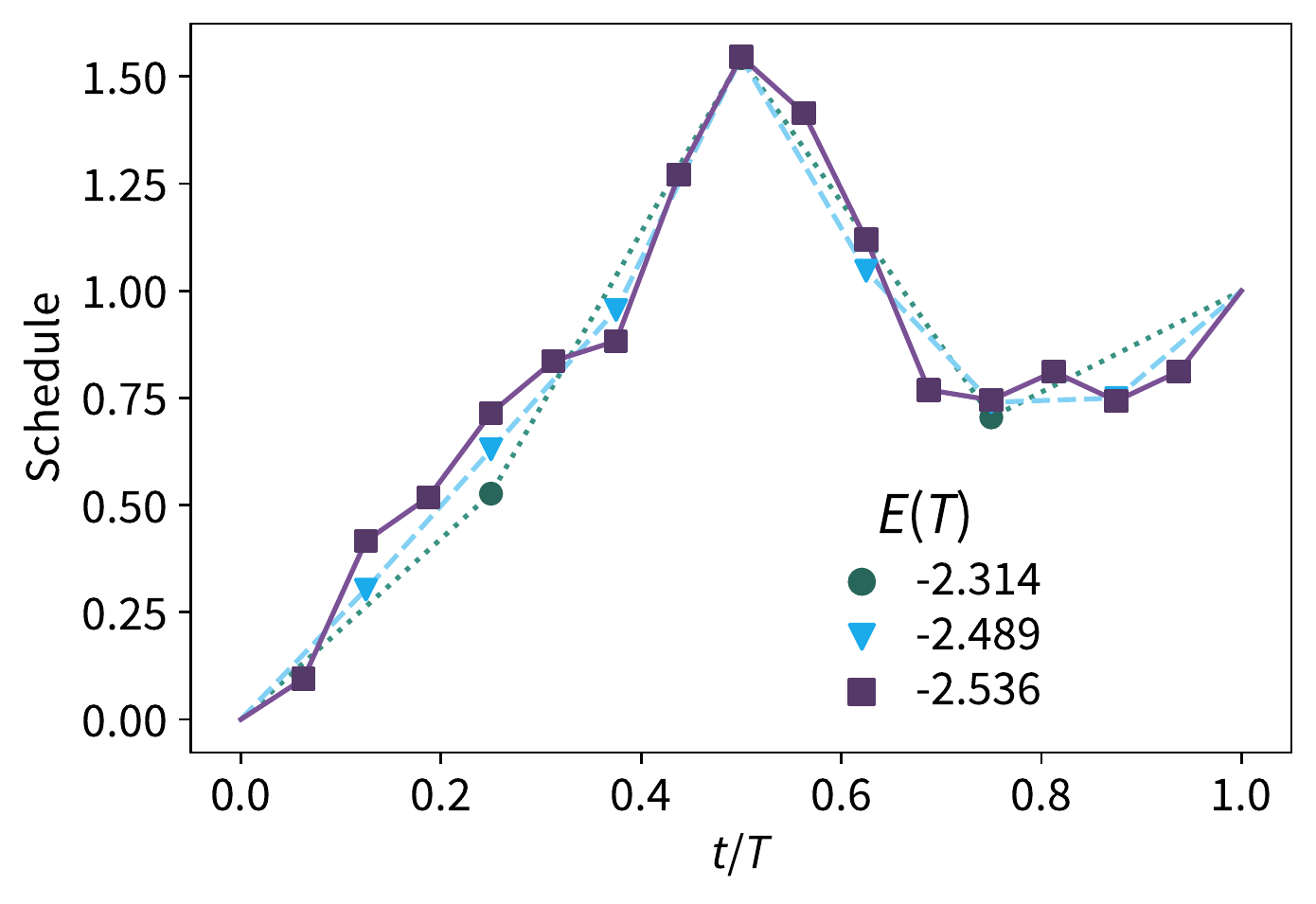}
    \caption{
    An example annealing schedule $A(t)$ for $N = 5$ during three stages of the optimization process, for $T = 12.5$. The green dotted line with circular markers is the initial guess with 3 points, the blue dashed line with triangular markers is the first optimized schedule with 7 points, and the purple solid line with square markers is the final optimized schedule with 15 points. The ground state energy is $E_{0} = -2.55$, and only the purple schedule reaches it within a tolerance of $c = 0.5$. The tolerance $c$ is defined in Eq.~\eqref{eq:EnergyThreshold}.
    }
    \label{fig:exampleSchedule}
\end{figure}

After initializing a random schedule, we optimize the parameters $a_j$ such that we minimize the energy of the output state $\ket{\psi(T)}$. We perform the optimization using a gradient-free method from SciPy, which we detail further in Appendix~\ref{sec:Hyperparameters}. The result is a locally optimal schedule $A(t)$ using a piece-wise linear function with three segments. In our experiments, we use ten different random instances of the triple $(a_1, a_2, a_3)$ and select the best set of parameters (in terms of the energy achieved) when proceeding to the next step.

Then, we refine the schedule by adding new points (i.e., additional free parameters) in between the original $(0,0)$, $(t_1, a_1)$, $(t_2, a_2)$, $(t_3, a_3)$ and $(T,1)$ points. In this case, we would go from three free parameters to seven (including the original three). In general, when we split the interval $\left[0, T \right]$ with $k$ points (not including the fixed endpoints), the points are located at $t_j = j T / (k+1)$, with $j \in \left\{ 1,2,\ldots,k \right\}$. We set the new values of $a_j$ to be along the line connecting the original points. This ensures that the new points do not change the schedule. This way, the optimization performed on a refined schedule is guaranteed to either improve or maintain the quality of the result (measured in energy $E(T)$).

For each new set of values $\left\{a_j \right\}$ defining $A(t)$, we run our optimization routine and compare the lowest energy found to the previous lowest energy value. If the lowest energy found does not decrease by more than a threshold value $\delta E$, we declare the energy ``converged'' and output $A(t)$ as our optimal annealing schedule (Fig.~\ref{fig:optimizeSchedule}). Otherwise, we add more points in between the current ones and optimize again. In Fig.~\ref{fig:exampleSchedule}, we show these intermediate steps of schedule refinement, from $3$ parameters to $15$ parameters that are required to converge close enough to the desired ground state.

In our optimization approach, we place no restrictions on the range of $A(t)$, but the schedules we find have a bounded range. This is important since in any physical implementation of the time-dependent Hamiltonian, the coupling strengths made available by the hardware will be limited. Furthermore, while the schedule's refinement was discussed using $3$ initial parameters, we also experiment with a number of $5$, $7$ and $9$ initial points. The results presented later on in Sec.~\ref{sec:Results}, are averaged over these different choices such that they are not tied to a specific choice of number of initial parameters. 

\subsection{Optimizing the annealing time}
\label{sec:OptimizingAnnealingTime}

\begin{figure}[ht]
    \centering    \includegraphics[width=0.48\textwidth]{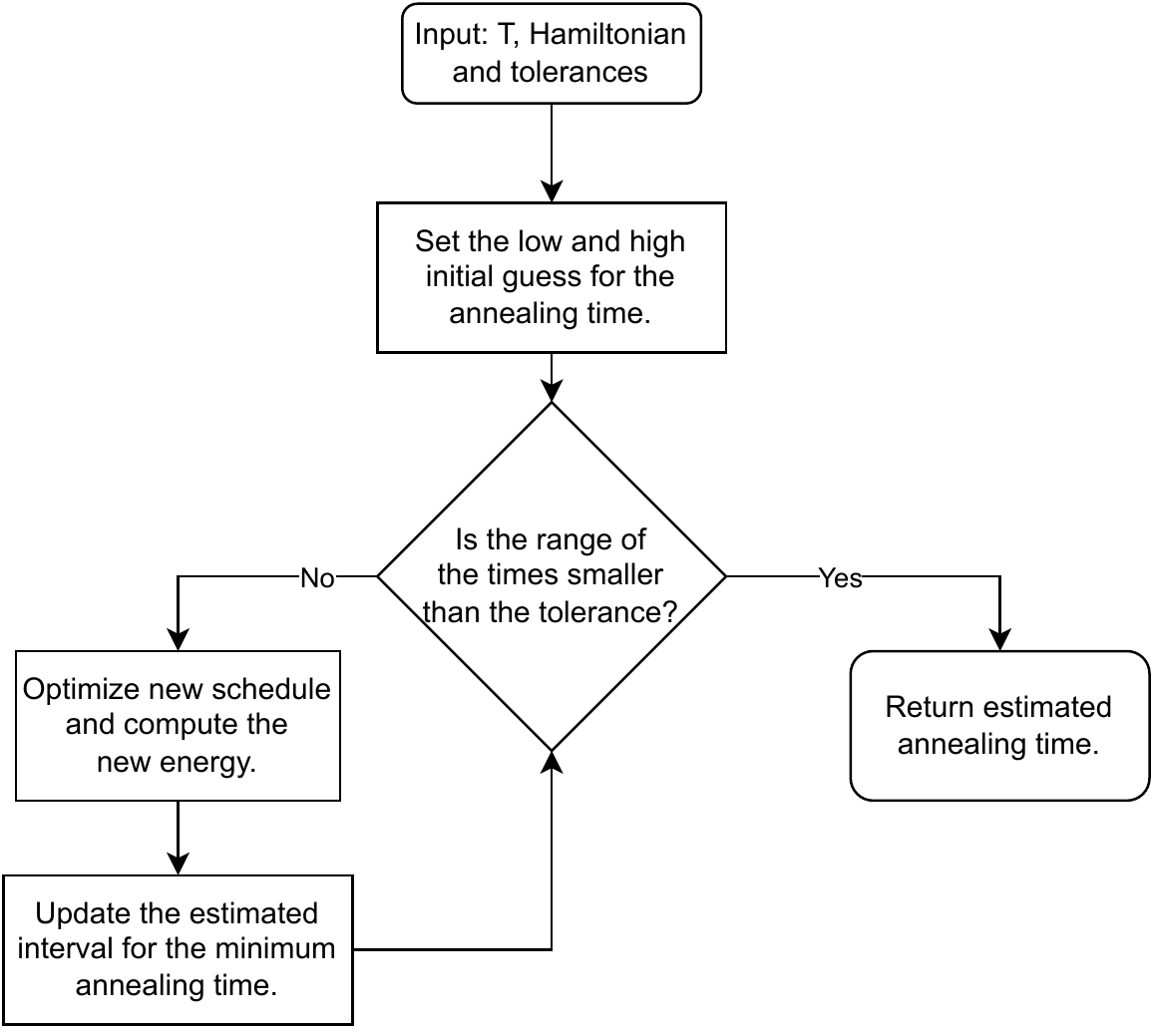}
    \caption{The outer routine of our optimization algorithm for optimizing the annealing time $T$. It provides our reasoning behind Algorithm \ref{alg:OptimizeT}.}
    \label{fig:optimizeT}
\end{figure}

The outer loop of the algorithm aims at determining the minimum annealing time $T_{\mathrm{min}}$ that keeps us within our threshold $E(T) - E_{0} \leq \Delta_E$ (see Fig.~\ref{fig:optimizeT} and Algorithm~\ref{alg:OptimizeT}).
We choose $\Delta_E$ to be a fraction $c$ of the problem Hamiltonian gap (Eq.~\eqref{eq:ExactExcitedEnergy}):
\begin{equation}
    \label{eq:EnergyThreshold}
    \Delta_E(c) = 2c \left(J_L - J_R \right), \,\,\, c > 0.
\end{equation}
When $c = 1$, the threshold corresponds to the gap between the ground state and the first excited state of the problem Hamiltonian. In practice, we vary $c$ to determine the performance of our algorithm. 

To estimate $T_{\mathrm{min}}$, we build an estimation interval $\left[ T_{\mathrm{low}}, T_{\mathrm{high}} \right]$ and take the estimate to be the upper limit of this interval. We start by setting $T_{\mathrm{low}} = 0$. To set an initial $T_{\mathrm{high}}$, we find a time $T$ such that the corresponding energy $E(T)$ after optimizing the schedule satisfies $E(T) - E_{0} \leq \Delta_E$. We declare such an optimization successful and set our current upper bound for the minimal annealing time as $T_{\mathrm{high}} = T$. Then, we update $T$:

\begin{equation}
    \label{eq:Tupdate}
    T \leftarrow \frac{T_{\mathrm{high}} + T_{\mathrm{low}}}{2}.
\end{equation}

As we continue refining our best estimate for $T_{\mathrm{min}}$, our interval of uncertainty $\left[ T_{\mathrm{low}}, T_{\mathrm{high}} \right]$ will shrink. We stop our algorithm when
\begin{equation}
    \label{eq:StoppingCondition}
    \frac{T_{\mathrm{high}} - T_{\mathrm{low}}}{T_{\mathrm{high}} + T_{\mathrm{low}}} \leq \delta T =  0.2,
\end{equation}
which is just the ratio between the uncertainty of the interval (half the range) and its midpoint $\left(T_{\mathrm{high}} + T_{\mathrm{low}} \right) / 2$.

By calculating $T_{\mathrm{min}}$ for multiple system sizes, we can study how the minimal annealing time scales with the system size. This is only an upper bound for the minimal annealing time $T_{\mathrm{min}}(N)$ satisfying $E(T) \leq E_0 + \Delta_E$, because our optimization algorithm could always fail to find a schedule with a smaller $T_{\mathrm{min}}$.

\begin{algorithm2e}
\caption{MinimizeAnnealingTime} \label{alg:OptimizeT}
\KwIn{$J_R$, $J_L$, $J$, $E_{0}$, $N$, $T$, $\delta T$, $\delta E, c$}
\KwOut{$T_{\mathrm{min}}$}
 $T_{\mathrm{low}} \gets 0$\;
 $T_{\mathrm{high}} \gets T$\;
 \While{$\frac{T_{\mathrm{high}} - T_{\mathrm{low}}}{T_{\mathrm{high}} + T_{\mathrm{low}}} \geq \delta T$ or $T_{\mathrm{high}} = T_{\mathrm{low}}$}{
    $A(t) \gets \mathrm{FindSchedule}$ \tcp*[h]{Use Algorithm~\ref{alg:OptimizeA}}\;
    $E(T) \gets \mathrm{ComputeEnergy}(A(t))$\;
    \eIf{$E(T) - E_{0} \leq 2c(J_L - J_R)$}{
        $T_{\mathrm{high}} \gets T$\;
    }{
        $T_{\mathrm{low}} \gets T$\;
    }
    $T \gets \frac{T_{\mathrm{high}} + T_{\mathrm{low}}}{2}$\;
}
$T_{\mathrm{min}} \gets T_{\mathrm{high}}$\;
\Return $T_{\mathrm{min}}$\;
\end{algorithm2e}

\section{Results}
\label{sec:Results}

In this section, we first compare the minimum annealing times obtained by means of our optimization (as described in Sec.~\ref{sec:methods}) to the times for a linear schedule to reach the same energy $E(T) \leq E_0 + \Delta_E$. We show that the optimized schedules can avoid the exponential times required by a linear schedule, resulting in annealing times scaling only polynomially with the system size. Second, we investigate what distinguishes the optimized schedules from the linear ones by studying the overlap of the instantaneous state with the ground state and first excited state of the instantaneous Hamiltonian $\hat{H}(t)$. In all the results we present, we set the system parameters to $\left(J_R, J_L, J \right) = \left(0.45, 0.5, 1 \right)$.

\subsection{Scaling of annealing time}
\label{sec:ScalingAnnealingTime}

\begin{figure}[ht]
    \centering
    \includegraphics[width=0.45\textwidth]{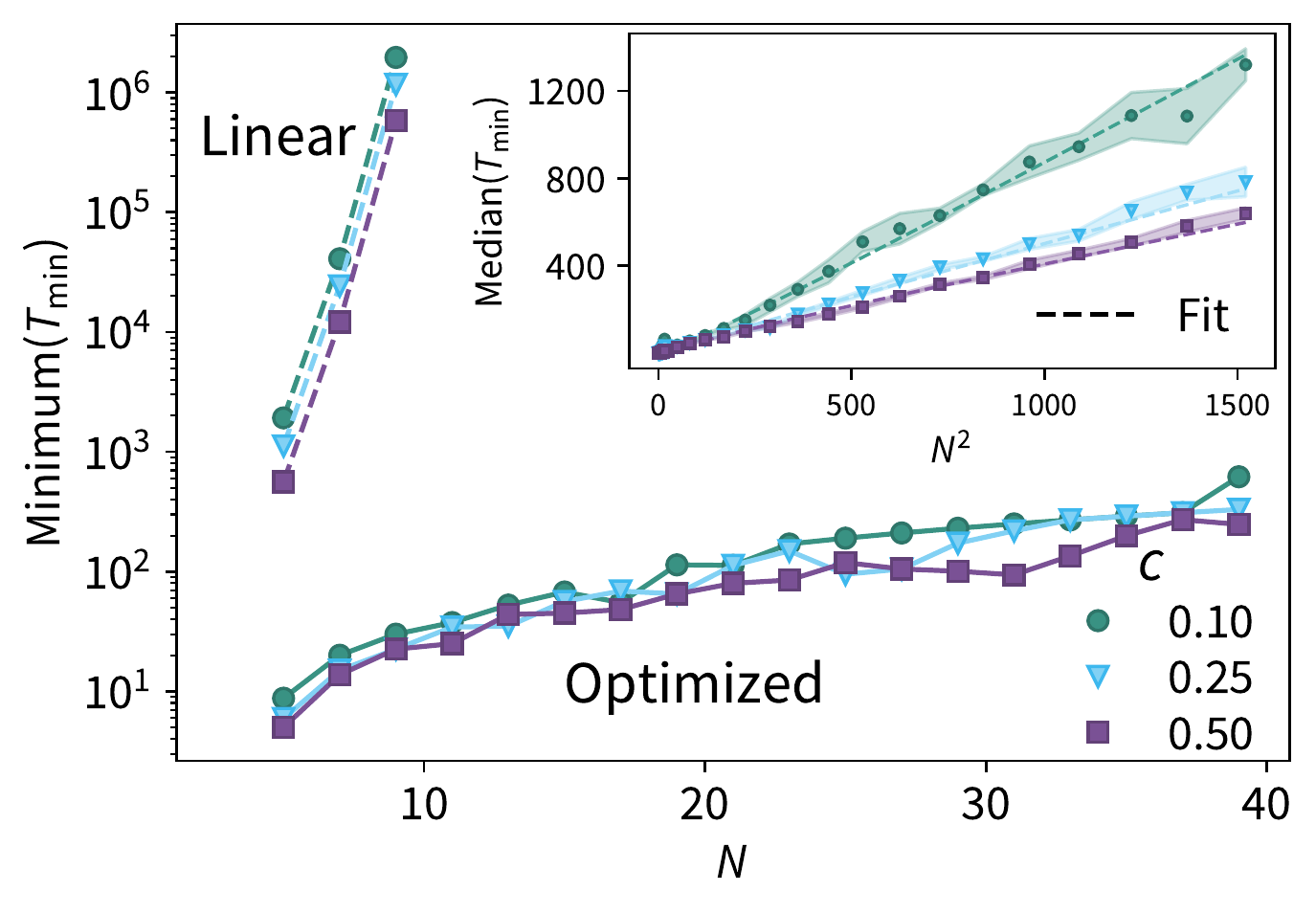}
    \caption{
    Scaling of the minimum annealing times $T_{\mathrm{min}}$ found for the optimized schedules (solid lines) and the linear ones (dashed lines). These are obtained for odd system sizes $N\in[5,39]$ and for different values of the energy threshold $c$ (colors in legend). Each data point in the ``Optimized'' part of the main figure represents the minimum annealing time found over $100$ runs of our algorithm. Furthermore, in the inset we report the median values of these times over the $100$ repetitions, along with a $95\%$ confidence interval in these estimates (shaded region). In both cases, we see evidence of the polynomial scaling of the optimized $T_{\mathrm{min}}$ with respect to $N$. In particular, the median values are found to scale quadratically with $N$, with a fit of the form $\alpha N^2$ as a leading term, depicted in the inset (dashed lines).
    }
    \label{fig:scalingAnnealingTime}
\end{figure}

In Fig.~\ref{fig:scalingAnnealingTime}, we report the annealing times corresponding to both the linear schedules and the optimized ones,  over different values of the threshold factor $c=0.1$, $0.25$ and $0.5$ appearing in Eq.~\eqref{eq:EnergyThreshold} (colors in legend). As expected from the discussion in Sec.~\ref{sec:Model}, we find that the linear schedules result in annealing times that grow exponentially fast with the size $N$ of the system. Already, for $N>9$ this would entail times $T > 10^6$ that are too computationally demanding to simulate.

In stark contrast, Fig.~\ref{fig:scalingAnnealingTime} demonstrates that our optimization protocol is capable of finding schedules significantly shorter than the linear ones (up to $5$ orders of magnitude shorter for $N=9$). 
Here, each data point corresponding to the optimized schedules is obtained as the minimum time found over $100$ optimizations. Overall, we see that  times only scale polynomially with $N$ for all the values of $c \in \left\{0.1, 0.25, 0.5 \right\}$ we study.

While this already provides evidence that optimized schedules can be exponentially shorter than linear ones, it is also of great interest to understand the scaling of annealing times entailed by a \emph{typical} optimized schedule. For that purpose, we report in the inset of Fig.~\ref{fig:scalingAnnealingTime}, the median values of the times found over the $100$ repetitions. As shown, these median times can be up to one order of magnitude larger than the minimum times found. Despite this increase, these median times still scale polynomially in $N$. Fits of the form $\mathrm{median}(T_{\mathrm{min}}) \sim \alpha(c) N^2$ (as a leading term) are displayed as dashed lines in the inset. As $c$ decreases, the slopes $\alpha(c)$ increases. 

\subsection{Population levels}
\label{sec:PopulationLevels}
\begin{figure}[ht]
    \centering
    \includegraphics[width=0.48\textwidth]{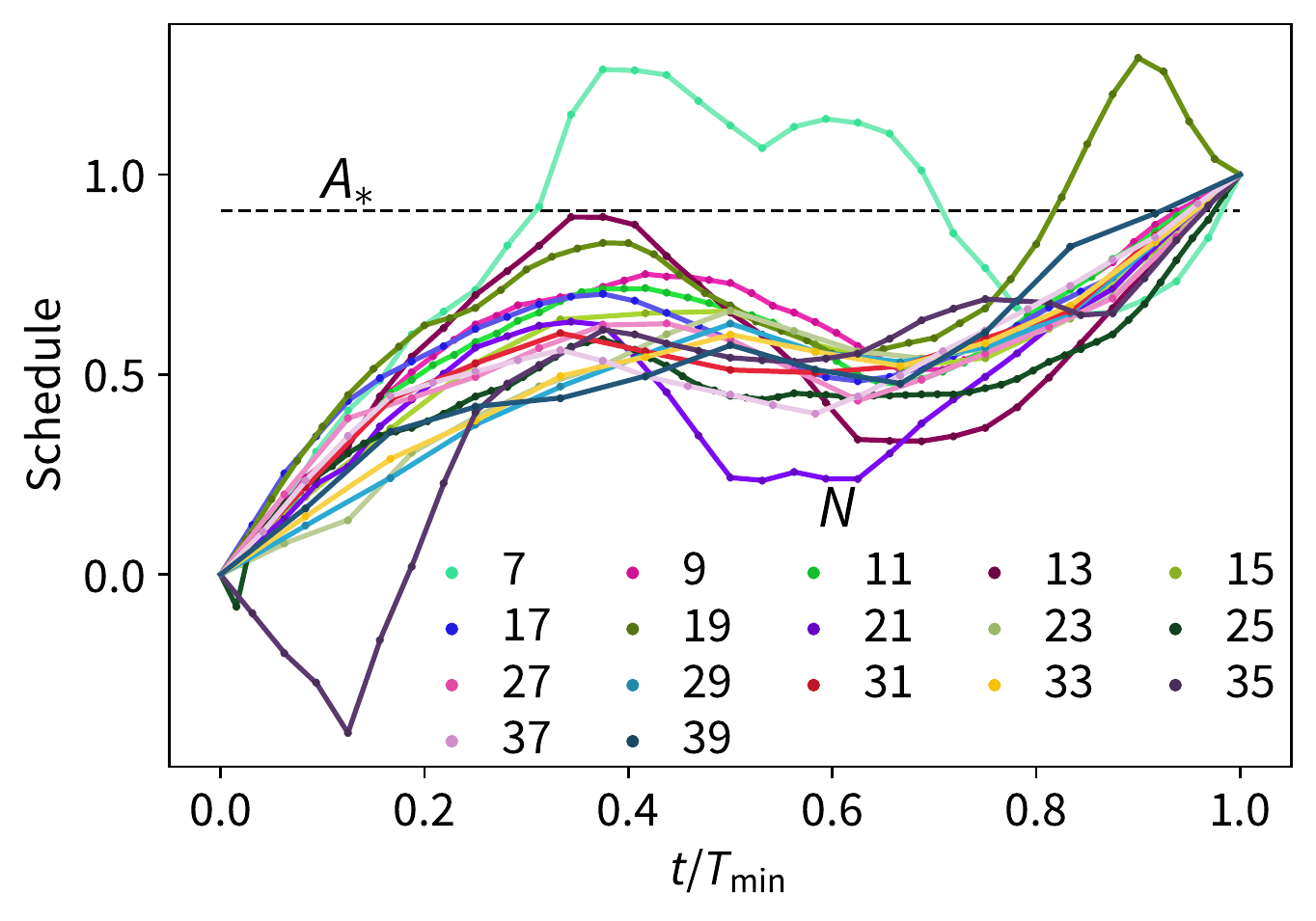}
    \caption{
    A selection of optimized schedules $A(t)$ across many system sizes that have been found for the case $c = 0.1$ in Eq.~\eqref{eq:EnergyThreshold}. 
    We plot each schedule along the time $t / T_{\mathrm{min}}$ that has been scaled by the overall time $T_{\mathrm{min}}$ of the schedule. Dots in the figure depict the optimized points for the schedule, with the initial and final ones fixed to $(0,0)$ and $(1,1)$. The dashed black horizontal line represents the value of schedule $A_\ast$ where there is a perturbative gap.
    }
    \label{fig:optimizedSchedules}
\end{figure}

To understand the underlying dynamics associated with our optimized annealing schedules and low annealing times, we first inspect the schedules themselves. In Fig.~\ref{fig:optimizedSchedules}, we report a selection of representative optimized schedules, identified for the threshold value $c=0.1$. These schedules share a common structure across various system sizes, in particular a similar ``up-down-up'' trajectory. 

To understand the role of the non-monotonic schedule, we look at the populations in the energy levels of the instantaneous Hamiltonian $\hat{H}'(t) = \hat{P}_N \hat{H}(t) \hat{P}_N$, where $\hat{P}_N = \left( \mathbb{\hat{1}}_N + \hat{X}^{\otimes N} \right) / 2$, and $\mathbb{\hat{1}}_N$ is the identity operator of dimension $2^N$. Because the initial state satisfies $\hat{P}_N \ket{\psi(0)} = \ket{\psi(0)}$ and our evolution preserves the symmetry ($[\hat{H}(t), \hat{X}^{\otimes N} ] = 0$), we can restrict ourselves to states in that symmetry subspace. This means that the relevant eigenvalues and eigenvectors are those of the Hamiltonian $\hat{H}'(t)$.

\begin{figure*}[ht]
    \centering
    \subfigure[$N=9$]{\includegraphics[width=1.0\textwidth]{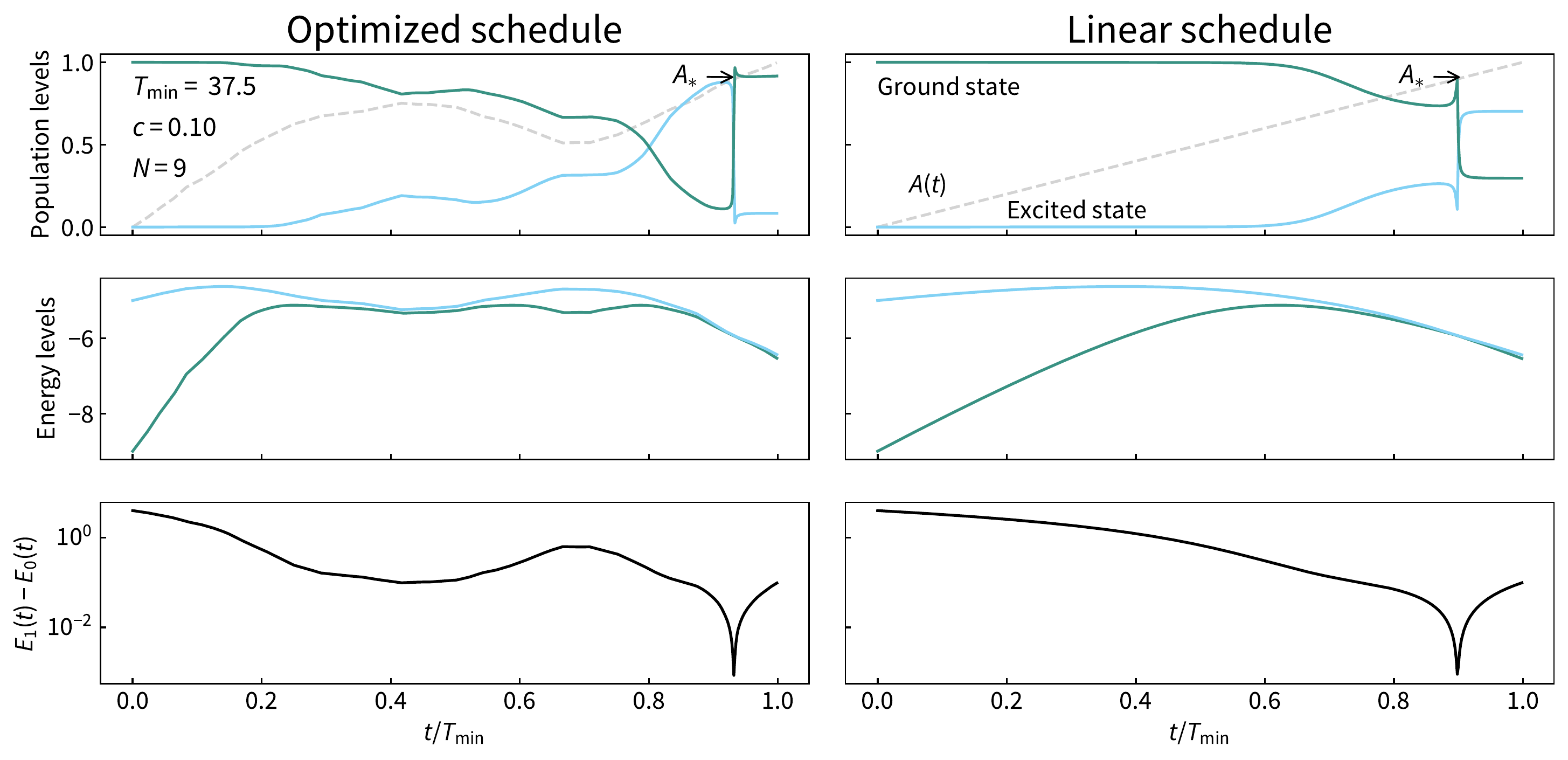}}
    \subfigure[$N=13$]{\includegraphics[width=1.0\textwidth]{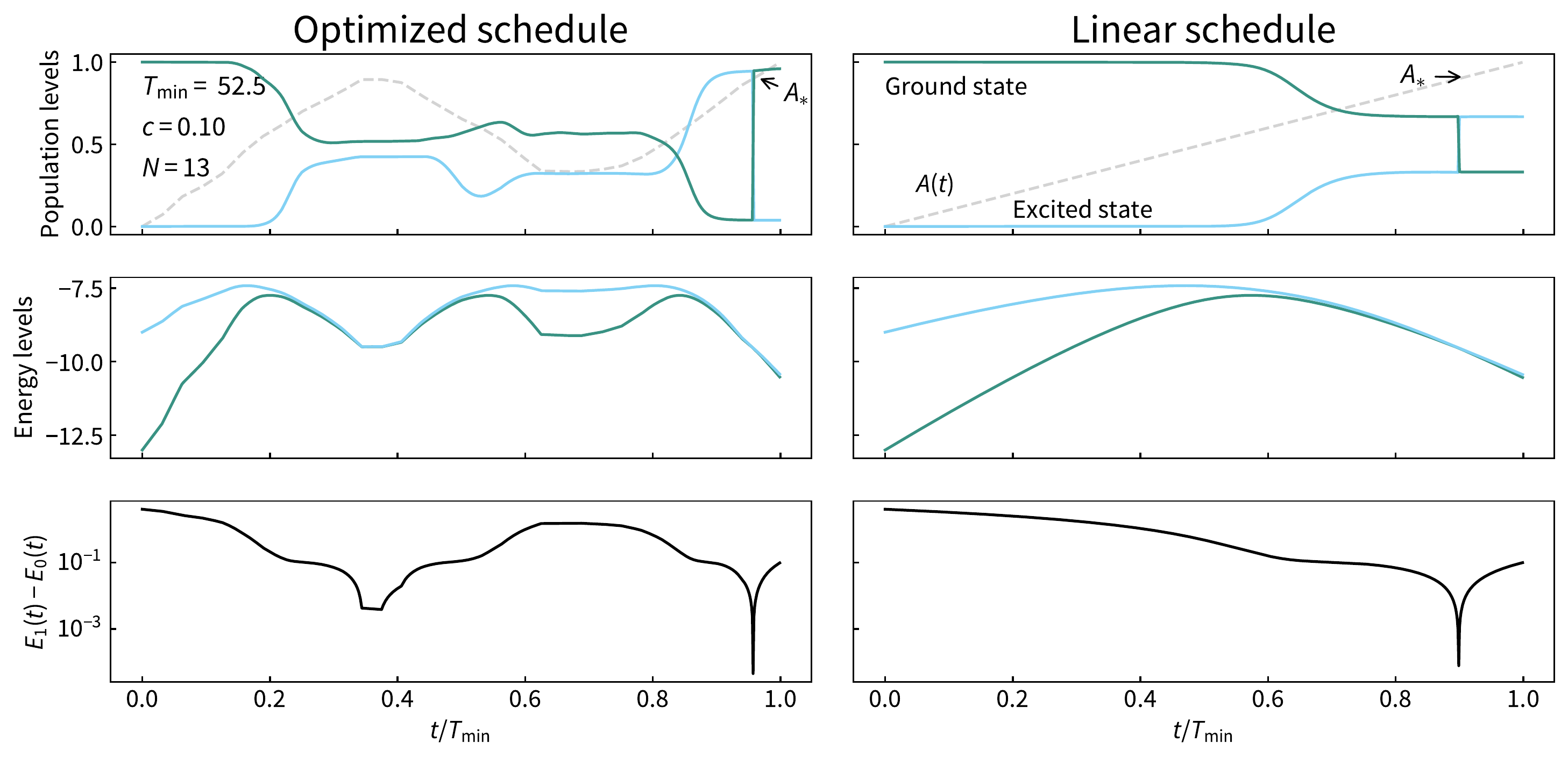}}
    \caption{Comparison of the linear versus optimized schedule for (a) $N=9$ and (b) $N=13$. Top row: The population levels of the ground and first excited state invert twice for the optimized schedule, versus only once in the linear case. Middle and bottom rows: This is because the energy gap between the two levels only closes at the perturbative crossing of $A_\ast \approx 0.9091$, which inverts the population levels in an undesirable way. On the other hand, the gap closes more than once for the optimized schedule, which allows the population levels to accumulate in the first excited state, and then transfer into the ground state when the system passes through $A_\ast$.}
    \label{fig:NPopulations}
\end{figure*}

By tracking how the populations in the energy levels of $\hat{H}'(t)$ change over time, we can observe how $A(t)$ affects the quantum state. Our simulations show that the majority of the energy population is restricted to the ground state and first excited state of $\hat{H}'(t)$. We show two examples in Fig.~\ref{fig:NPopulations}. The strategy followed by the optimized schedule is to first transfer population from the ground state to the first excited state and then cross the perturbative gap at $A \approx A_\ast$ one last time to diabatically transfer the population back into the ground state~\cite{Cro2014}. What is of particular interest is that the transfer of population to the first excited state is done by using a non-monotonic schedule. In some cases, the function $A(t)$ crosses the perturbative gap at $A_\ast$ multiple times. For example, the schedule in Fig.~\ref{fig:exampleSchedule} crosses the value $A_\ast \approx 0.9091$ a total of three times (twice increasing above $A_\ast$ and once decreasing below it). In others, the function $A(t)$ approaches the perturbative gap at $A_\ast$ but does not cross it, moves away from it, and only crosses it near the end of the evolution. We show this in  Fig.~\ref{fig:optimizedSchedules}, where the horizontal dashed line indicates the schedule value $A_\ast$ which leads to the perturbative gap.

To put our results into context, we compare our optimized schedules to linear schedules for the same optimized time $T_{\mathrm{min}}$ in Fig.~\ref{fig:NPopulations}. Doing so highlights the advantages of a non-adiabatic evolution. We look at the instantaneous population levels, energy levels, and gap to the ground state along both schedules. We observe that the linear schedule only creates a small gap near the exponential closing $A_\ast$. If we anneal too fast, this exponentially small gap will cause the population levels of the ground and first excited state to invert. The only option is to anneal more slowly, which lowers the probability that the population levels invert. By contrast, some of our optimized schedules first make the gap to the ground state small \emph{without} crossing $A_\ast$, allowing most of the population to migrate from the ground state to the first excited state. Then, the quantum state anneals through the point $A_\ast$, inverting the populations so the ground population is the majority.

A common feature of our optimized schedules is the ``plateau'' in the populations at intermediate times. Naively, this might suggest that the annealing time could be further reduced since the system appears to be ``idle'' during the plateau period, but we believe that the relative phase accumulation during this period is crucial to get a high first excited population near the end of the anneal that then transitions to a high ground state population. 

In addition to gaining a better understanding of the physics at play, identifying common features of the optimized schedules as shown in Fig.~\ref{fig:optimizedSchedules} could serve to develop more informed initialization strategies, or to potentially refine the way the schedules are parameterized. For instance, one would expect that using the patterns identified so far as an initial guess for the schedules to be optimized over (rather than the random initialization adopted in this study) would significantly improve the convergence of our algorithm. We leave this study for future work.

\section{Conclusion}
\label{sec:Conclusion}

Our work shows that we can trade the exponential annealing time required using an adiabatic approach for a polynomial annealing time using an optimized schedule with a non-adibatic evolution. By examining the population levels of a few schedules, we showed that the non-adiabatic nature of the schedules helps the system anneal more quickly into the ground state. Our Ising Hamiltonian (Eq.~\eqref{eq:Hamiltonian}) is one dimensional, so from a classical optimization perspective it is easy to find the ground state using heuristic algorithms since growing the size of domain walls along the spin chain does not cost additional energy. The problem's exponential hardness for adiabatic quantum annealing stems from a perturbative crossing that arises from how the transverse field breaks the degeneracy of the low-lying energy eigenstates. Our work suggests that there are relatively simple annealing schedules that can allow for a polynomial time annealing algorithm to solve this problem. This highlights the importance of sufficient flexibility in the annealing schedule to achieve such an advantage.

However, it should be noted that the saving in annealing time could be offset by the effort spent in optimizing the schedules in the first place, rendering such an approach less attractive.  We provide several points of indirect evidence that the cost of optimization remains under control. Most importantly, for the fixed amount of computational resources allocated for each run of the optimization of the algorithm, we were able to optimize schedules for up to $N = 39$, which we would not expect if the optimization effort was growing exponentially with the system size. Furthermore, we fixed the amount of computational resources when running each schedule optimization and did not see any particular increase in the number of steps of schedules refinement required as the system size increases. This suggests that for this problem Hamiltonian, the cost of the schedule optimization is unlikely to be exponential in the system size. 

We can consider following a similar strategy for solving more complex problems. For spin glasses, we can expect our schedule optimization to help us reach the low energy states of the rugged landscape more quickly, but it is unrealistic to imagine that an efficient schedule optimization would allow us to identify the global optimum efficiently. Nevertheless, it may be sufficient to find high quality low-energy solutions fast depending on the optimization context.

We do not claim that the schedules we have identified are optimal in the sense that they achieve the highest ground state probability for the lowest annealing time. We have observed that different choices of initial schedules can affect our heuristic algorithm, so the optimization landscape may include schedules with different characteristics and of a higher quality than the ones we present. A thorough investigation of this possibility using quantum optimal control for quantum annealing~\cite{Bra2021} would however be limited to small system sizes due to the high computational cost.

It is worth noting that the performance of DQA is likely to be related to the performance of another quantum algorithm for tackling optimization problems, the Quantum Approximate Optimization Algorithm (QAOA)~\cite{Far2014},  in the quantum circuit paradigm. Recent works~\cite{Pag2020,Zho2020} have highlighted that in the heuristic setting where the QAOA parameters are trained, the algorithm can be understood as implementing a diabatic evolution as in DQA. In fact, it has been shown that a hybrid of the two approaches, whereby the unitary evolution is allowed to be defined by both a continuous interpolating Hamiltonian as well as Hamiltonians that are turned on and off, is in fact optimal for solving optimization problems in this heuristic way~\cite{Bra2021, Bra2021b}.  Our results for DQA may therefore indicate that similar behavior might be observed for QAOA.

As identified and discussed in Sec.~\ref{sec:PopulationLevels}, the existence of a shared structure of the optimized schedules across different system sizes could serve as a basis of improved initialization of the algorithm. Such feature has been observed recently in the context of QAOA~\cite{Zho2020,Pag2020,sauvage2021flip,mele2022avoiding}. Given the difficulties that would be encountered when performing optimization directly based on measurement data  - with in particular the flattening of the cost function values~\cite{mcclean2018barren,cerezo2020cost} - this may prove crucial for realistic optimization of annealing schedules at scale.

There are several aspects of our optimization protocol that could be modified. These include the choice of driver Hamiltonian or intermediate Hamiltonian \cite{Zen2016,Mat2021}, the initial quantum state, and the way we parameterize the schedule. An optimization procedure that includes the cost of identifying the optimal schedule using optimal stopping techniques~\cite{Vin2016} would verify the viability of such strategies for real world optimization.

Our hope is that such optimization protocols can be used to create targeted schedules for highly-controllable quantum annealers, offering non-trivial operating benchmarks for such devices as well as removing the bottlenecks of the standard adiabatic paradigm when possible.

\section*{Data Availability}
The data for the optimized schedules in Fig.~\ref{fig:optimizedSchedules}, the scaling of the annealing time, and the population levels is available at the following Zenodo repository: \url{https://doi.org/10.5281/zenodo.7392024}.

\acknowledgments{
This work was supported by the U.S. Department of Energy (DOE) through a quantum computing program sponsored by the Los Alamos National Laboratory (LANL) Information Science \& Technology Institute. L.C. and M.L. were supported by the U.S. DOE, Office of Science, Office of Advanced Scientific Computing Research, under the Accelerated Research in Quantum Computing (ARQC) program. F.S. acknowledges Laboratory Directed Research and Development (LDRD) program of LANL under project number 20220745ER. J.C. was supported by a B2X scholarship from the Fonds de recherche--Nature et technologies and a scholarship from the Natural Sciences and Engineering Research Council of Canada [funding reference number: 456431992]. J.C. also acknowledges the physics PhD program at the Universit\'{e} de Sherbrooke and funding from the Canada First Research Excellence Fund. This material is based upon work supported by the National Science Foundation under Grant No. 2037755. T.A. thanks Jarrett Smalley for useful discussions.
}

\bibliographystyle{apsrev4-2}
\bibliography{refs}

\appendix

\section{Simulating the schedule}
\label{sec:Simulation}

For the optimizations in Sec.~\ref{sec:ScalingAnnealingTime}, we study system sizes of up to $N=39$ qubits. Simulations with such a large number of qubits are possible by numerical routines first proposed in~\cite{somma2005quantum} and implemented in~\cite{goh2022atheory}. The simulation works at the level of the system's dynamical Lie algebra~\cite{larocca2021diagnosing,larocca2021theory}, the Lie closure of the set $\{i\hat{H}_d,i\hat{H}_p\}$, which in our case scales efficiently (polynomially) in the system size. We perform Heisenberg evolution of the observable $\hat{H}_p$. Crucially, irrespective of the time $T$ considered and irrespective of the schedule employed, the evolved operator remains contained in the dynamical Lie algebra, and therefore admits an efficient description. That is, by keeping track of the instantaneous support of the observable over a basis of the dynamical Lie algebra, one can assess the energy of the evolved system at scale.

However, the study of the instantaneous population in Sec.~\ref{sec:PopulationLevels} requires full knowledge of the system over time, and therefore we resort to conventional full-state vector simulations, thus limiting the number of qubits we can simulate. 
For the example schedule of Fig.~\ref{fig:exampleSchedule} and the population levels in Sec.~\ref{sec:PopulationLevels}, we use Qiskit~\cite{Qiskit} to simulate the time evolution of the schedule. We write the time-evolved state $\ket{\psi(T)}$ at the end of the annealing procedure as
\begin{equation}
    \label{eq:TimeEvolvedState}
    \ket{\psi (T)} = \hat{U}(T) \ket{\psi(0)},
\end{equation}
where $\ket{\psi(0)} = \ket{+}^{\otimes N}$, and the unitary $\hat{U}(T)$ is the time-ordered propagator
\begin{equation}
    \label{eq:Propagator}
    \hat{U}(T) \equiv  \mathrm{T}\exp{\left( - i \int_0^T \hat{H}(t) \mathrm{d}t \right)} .
\end{equation}
We approximate $\hat{U}(T)$ with small time slices $\delta t$ and implement the evolution as a quantum circuit. This means the unitary in Eq.~\eqref{eq:Propagator} decomposes as:
\begin{equation}
    \label{eq:DecomposedUnitary}
    \hat{U}(T) \approx U( T, T-\delta t) \dots  \hat{U}(2\delta t, \delta t) \hat{U}(\delta t, 0) ,
\end{equation}
where $\hat{U}(t + \delta t, t)$ is the unitary between times $t$ and $t + \delta t$. For small enough $\delta t$, we assume the Hamiltonian $\hat{H}(t)$ is constant in a given interval.

Let us focus our attention on one particular time slice at time $t^*$, with the associated unitary operator
\begin{equation}
\begin{aligned}
    \label{eq:UnitarySlice}
    \hat{U}(t^* + \delta t, t^*) &\equiv \exp{ - i \hat{H}(t^*) \delta t } \\ 
    &= \exp{- i \delta t \left( A(t^*) \hat{H}_p + (1 - A(t^*)) \hat{H}_d \right)}.
\end{aligned}
\end{equation}
Because the terms in $\hat{H}_d$ and $\hat{H}_p$ do not commute (one contains $\hat{Z}$ terms and the other contains $\hat{X}$ terms), we decompose the unitary slice with a first-order Trotter decomposition~\cite{Tro1959,Suz1976}:
\begin{equation}
\begin{aligned}
    \label{eq:TrotterApproximation}
    \hat{U}(t^* + \delta t, t^*) &\approx \exp{- i \delta t \left( - A(t^*) \sum_{j = 1}^N J_{j} \hat{Z}_j \hat{Z}_{j+1} \right)} \\ & \times \exp{- i \delta t \left(- (1-A(t^*)) \sum_{j = 1}^N \hat{X}_j \right)}.
\end{aligned}
\end{equation}
Within each exponential in Eq.~\eqref{eq:TrotterApproximation}, the operators commute with each other for all $i$, which allows us to rewrite the sum as a product. We then find the final form for our unitary:

\begin{equation}
\begin{aligned}
    \label{eq:TrotterGates}
    \hat{U}(t^* + \delta t, t^*) &\approx \prod_{j = 1}^N \exp{- i \delta t \left( - A(t^*) J_{j} \hat{Z}_j \hat{Z}_{j+1} \right)} \\ &\times \prod_{j = 1}^N \exp{- i \delta t \left(- (1-A(t^*)) \hat{X}_j \right)}.
\end{aligned}
\end{equation}

Equation~\eqref{eq:TrotterGates} tells us that each time slice of size $\delta t$ requires $2N$ gates (there are $N$ terms in each product), and the two types of quantum gates are:

\begin{equation}
    \label{eq:Gates}
    \begin{aligned}
        \mathrm{RZZ} \left( j, j+1, \theta_Z \right) &\equiv \exp{-i \frac{\theta_Z}{2} \hat{Z}_j \hat{Z}_{j+1} }, \\
        \mathrm{RX} \left(j, \theta_X \right) &\equiv \exp{-i \frac{\theta_X}{2} \hat{X}_j},
    \end{aligned}
\end{equation}
where in our case the angles are $\theta_Z = -2 \delta t A(t^*) J_j$ and $\theta_X = -2 \delta t (1-A(t^*))$.

\begin{figure}[ht]
    \centering
    \includegraphics[width=0.48\textwidth]{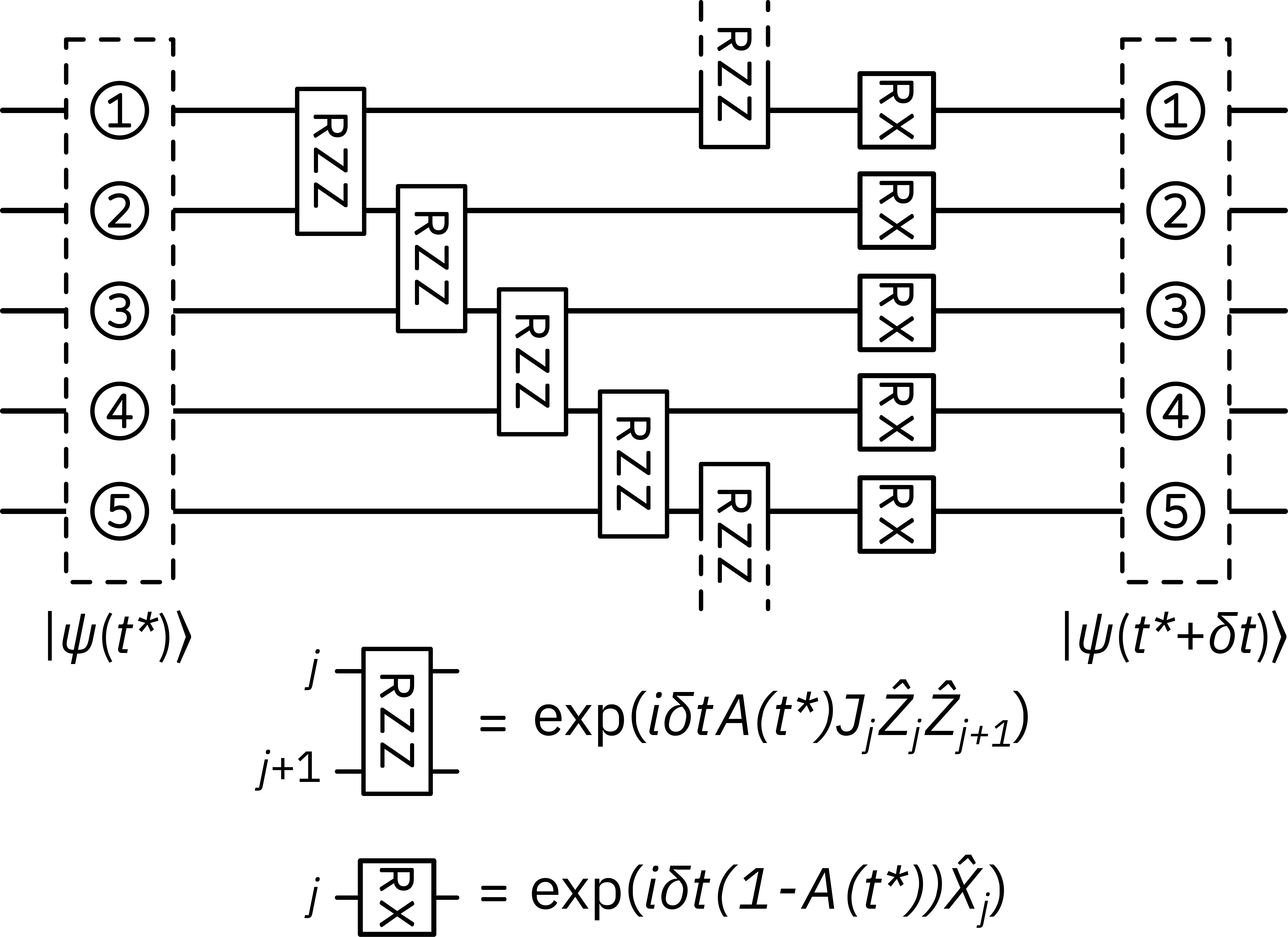}
    \caption{The quantum circuit for the evolution given by Eq.~\eqref{eq:TrotterGates} for $N = 5$. The circles label the physical sites of the ring, and the $\mathrm{RZZ}$ gate at the top and bottom connects sites $1$ and $N$. This is the circuit for one time slice between $t^*$ and $t^* + \delta t$.}
    \label{fig:circuit}
\end{figure}

This collection of gates in Eq.~\eqref{eq:TrotterGates} defines a quantum circuit for each time slice. The full evolution involves the same layout of gates over and over again, but with varying $\theta_Z$ and $\theta_X$, according to the schedule $A(t)$, and the couplings $J_j$. In Fig.~\ref{fig:circuit}, we show the circuit according to this layout.

In order to calculate $E(T) \equiv \expval{\hat{H}_p}{\psi(T)}$, we choose to discretize the unitary evolution operator in terms of time steps $\delta t$, and the output state is then given by $\ket{\psi(T, \delta t)}$. Calculating the energy then requires sending the discretization to zero:
\begin{equation}
    \label{eq:ConvergedEnergy}
    E(T) = \lim_{\delta t \rightarrow 0} \expval{\hat{H}_p}{\psi(T, \delta t)}.
\end{equation}
We do this by building the circuit for multiple values of $\delta t$, and checking how the energy $E(T)$ changes as we decrease the size of $\delta t$. We use the same threshold $\delta E$ as in Sec.~\ref{sec:OptimizingSchedule} to stop this procedure, giving us our ``converged'' estimate of $E(T)$ (see Appendix~\ref{sec:Hyperparameters}).

To calculate the quantities from Sec.~\ref{sec:PopulationLevels}, we track the instantaneous quantum state $\ket{\psi(t)}$ during the annealing schedule. We also calculate the instantaneous Hamiltonian in the relevant symmetry subspace 
$\hat{H}'(t) = \hat{P}_N \hat{H}(t) \hat{P}_N$, where $\hat{P}_N = \left( \mathbb{\hat{1}}_N + \hat{X}^{\otimes N} \right) / 2$, and $\mathbb{\hat{1}}_N$ is the identity operator of dimension $2^N$. Diagonalizing $\hat{H}'(t)$ provides us with $E_0(t)$ and $E_1(t)$ in the relevant symmetry subspace, so we can calculate the relevant energy gap for our dynamics.

In order to calculate the instantaneous populations of a given energy eigenstate, we find the eigenvectors $\ket{E_k(t)}$ of $\hat{H}'(t)$, with $k$ indexing the energy levels. We then define the population level as:

\begin{equation}
     \label{eq:PopulationLevels}
     P_k(t) \equiv \rvert \braket{\psi(t)}{E_k(t)} \lvert^2, \,\,\, k \in \left\{0, 1, 2, \ldots, 2^N-1 \right\},
 \end{equation}
where $\ket{E_k(t)}$ is an equal superposition of eigenstates of the Hamiltonian $\hat{H}'(t)$ with energy $E_k(t)$. Note that for most cases, there is a unique state for each energy level. The population levels are fractions, so they must sum to one:

\begin{equation}
     \label{eq:PopulationSum}
     \sum_{k = 0}^{2^{N}-1} P_k(t) = 1.
 \end{equation}
 
In Figs.~\ref{fig:NPopulations}, we show $P_0(t)$ and $P_1(t)$, since they contribute the most to the annealing dynamics.

\section{Implementation details}
\label{sec:Hyperparameters}

For our numerical work, we use the minimization procedure implemented in SciPy with the COBYLA method~\cite{powell1994direct}. We set the maximum number of iterations ``maxiter'' of the function to be $800$, and the tolerance for the gradient ``tol'' to be $0.001$.

For every numerical experiment, we set $J_R = 0.45$, $J_L = 0.50$, and $J = 1$. We set the initial $\delta t$ to be $1$.

The thresholds we use to determine when to stop our algorithm are:

\begin{enumerate}
    \item $\delta T = 0.1$
    \item $c \in \left\{0.1, 0.25, 0.50 \right\}$
    \item $\delta E = 0.001$ 
\end{enumerate}

\end{document}